\documentstyle[times,epsf]{mn}

\title[X-ray jets common in low-power radio galaxies]{{\it Chandra\/}
finds X-ray jets are common in low-power radio galaxies}

\author[D.M. Worrall et al.]{D.M.~Worrall, M.~Birkinshaw and 
M.J.~Hardcastle\\
Department of Physics, University of Bristol, Tyndall Avenue,
Bristol BS8 1TL}

\begin{document}

\maketitle

\label{firstpage}

\begin{abstract}

We present results for the first three low-power radio galaxies from
the B2 bright sample to have been observed with {\it Chandra\/}.  Two
have kpc-scale radio jets, and in both {\it Chandra\/} resolves jet
X-ray emission, and detects soft X-ray core emission and an X-ray
emitting galaxy-scale atmosphere of luminosity a few $10^{41}$ ergs
s$^{-1}$. These are the first detections of X-ray jets in low-power
radio galaxies more distant than Cen~A and M~87. The cooling time of
the galaxy-scale gas implies mass infall rates of order one solar mass
per year. The gas pressure near the jets is comparable to the minimum
pressure in the jets, implying that the X-ray emitting gas may play an
important role in jet dynamics.  The third B2 radio galaxy
has no kpc-scale radio jet, and here only soft X-ray emission from the
core is detected.  The ratio of X-ray to radio flux is similar
for the jets and
cores, and the results favour a synchrotron origin for the
emission. kpc-scale radio jets are detected in the X-ray in $\sim
7$~ks exposures with {\it Chandra\/} more readily than in the optical
via {\it HST\/} snapshot surveys.

\end{abstract}


\section{Introduction}

Radio observations of low-power FRI (Fanaroff \& Riley 1974) radio
galaxies are best interpreted by models in which the jet plasma slows
from relativistic ($v \sim 0.9c$) to sub-relativistic ($v \leq 0.1c$)
speeds within a few kpc of the nucleus (Laing et al.~1999).  Decreasing
jet asymmetries at sub-kpc to kpc distances from the nucleus can then
be understood as the diminishing effect of Doppler boosting. The jets
at termination blend into the lobe plasma they supply.  In contrast,
the jet asymmetry seen in powerful, FRII, sources supports velocities
as large as $\sim 0.7c$ at tens of kpc or more
from their cores (Wardle \& Aaron 1997).  Bright hotspots mark
jet termination in these objects.

The relativistic particles responsible for the jets in FRI and FRII
sources are expected to emit elsewhere in the electromagnetic
spectrum, and the X-ray is no exception.  However, it is jet features
of the powerful FRII sources which have so far yielded the largest
number of X-ray detections.  These are (a) hotspots in Cyg~A (Carilli,
Perley and Harris 1994), 3C~295 (Harris et al.~2000), and 3C~123
(Hardcastle, Birkinshaw \& Worrall 2001a), where the X-rays have been
interpreted as due to synchrotron self-Compton (SSC) emission of
plasma at minimum energy, (b) hotspots in 3C~390.3 (Prieto 1997),
3C~120 (Harris et al.~1999), and Pic~A (Wilson, Young \& Shopbell
2001), where the interpretation is less clear, and (c) emission
associated with large-scale jets in 3C~273 (Marshall et al.~2001),
PKS~0637-752 (Schwartz et al.~2000), and Pic~A (Wilson et al.~2001).

Low-power radio galaxies typically reside in group-scale X-ray
emitting atmospheres (Worrall \& Birkinshaw 2000).  While the
evacuation of the hot X-ray plasma from the radio lobes has apparently
been seen in sources such as NGC 1275 (B\"ohringer et al.~1993),
3C~449 (Hardcastle, Worrall \& Birkinshaw~1998), and M~84 (Finoguenov
\& Jones~2001), and evidence for inverse Compton X-rays from the lobes
in Fornax~A (Feigelson et al.~1995), jet X-ray emission has been
reported only in the two nearby low-power sources Cen~A
(D\"obereiner et al.~1996; Kraft et al.~2000) and M~87 (Biretta, Stern
\& Harris~1991; B\"ohringer et al.~2001).

\begin{table*}
\caption{B2 sources observed with {\it Chandra}}
\label{xrayobs}
\begin{tabular}{llrccllclc}
Name &
$z$ &
408 MHz power &
kpc/ &
Galactic N$_{\rm{H}}$ &
\multicolumn{2}{c}{J2000 X-ray core position} &
shift &
Date  & 
Screened 
\\
& & W Hz$^{-1}$ sr$^{-1}$ &
arcsec & ($10^{20}$~cm$^{-2}$) & \multicolumn{2}{c}{J2000 radio core position} 
&  (arcsec) 
& & Exposure (ks) \\
B2 0206+35 & 0.0369 & 2.42 $\times 10^{24}$ 
 & 1.02 & 5.9 & 02 09 38.57 & 
$+$35 47 48.3 & 2.6 & 2000 Mar 18 & 7.387\\
&&&&& 02 09 38.560 & $+$35 47 50.92 &&& \\
B2 0331+39 & 0.0204 & 0.261 $\times 10^{24}$ & 
0.58 & 14.6 & 03 34 18.40 & 
$+$39 21 22.2 & 2.3 & 2000 Apr 1 & 5.150\\
&&&&& 03 34 18.419 & $+$39 21 24.44 &&& \\
B2 0755+37, NGC 2484 & 0.0428 & 3.80 $\times 10^{24}$ 
& 1.17 & 5.02 & 
07 58 27.95 & $+$37 47 10.6 & 2.2 & 2000 Apr 3 & 6.716\\
&&&&& 07 58 28.108  &  $+$37 47
11.81 &&& \\
\end{tabular}
\medskip
\begin{minipage}{\linewidth}
We use $H_o = 50$ km s$^{-1}$ Mpc$^{-1}$, $q_o = 0$, throughout.
Uncertainties in positions of the radio cores, from data of Figs. 1-3,
are estimated to
be $\pm 0.05$ arcsec for B2~0206+35 and 0331+39 and
$\pm 0.03$ arcsec for B2~0755+37.  Redshift and radio power from
Colla et al.~(1975).
\end{minipage}
\end{table*}

We have embarked on a high spatial-resolution study with {\it
Chandra\/} of the X-ray emission from the inner regions of typical
low-power radio galaxies, where the radio-emitting plasma is expected
to have relativistic bulk motion.  Our targets are drawn from the
bright B2 sample, which is the complete subset of 50 sources from the
B2 radio survey identified with elliptical galaxies brighter than
$m_{\rm ph} = 15.7$ mag and well matched to BL~Lac objects in extended
radio properties and galaxy magnitudes (Colla et al.~1975; Ulrich
1989).  {\it ROSAT\/} pointed observations with either the PSPC or HRI
(or in some cases both) have been made for 80 per cent of the sources
in the bright B2 sample (Canosa et al.~1999), making this the largest
unbiased sample of exclusively low-power radio galaxies with sensitive
X-ray data.  We used a correlation of core unresolved (to {\it
ROSAT\/}) X-ray and radio emission to argue for a component of soft
X-ray emission associated with the inner radio jets (Canosa et
al.~1999).  However, the spatial resolution and sensitivity were
insufficient to distinguish between truly small-scale non-thermal
emission and small-scale hot gas possibly associated with cooling
flows, and most observations were with the {\it ROSAT\/} HRI with
which no spectral information was available.  Here we present results
for the first three B2 sources to have been observed with {\it
Chandra\/}.  They indicate that resolvable X-ray jets are common even
in low-power radio galaxies.

\section{{\it Chandra\/} Observations}

We observed B2~0206+35, 0331+39 and 0755+37 with the Advanced CCD
Imaging Spectrometer (ACIS) on board the {\it Chandra X-ray
Observatory\/} on the dates given in Table~\ref{xrayobs}.  The targets
were near the aim point on the back-illuminated CCD chip S3, 
shifted slightly from the original aim point to avoid sitting on
a boundary between readout nodes. The
observations were made in window mode with frame times of 0.44~s for
B2~0331+39 and 0.87~s for the other two sources, in order to guard
against the effects of pile-up should all the flux which was
unresolved to {\it ROSAT\/} be unresolved also to {\it Chandra\/}.
The data provided to us had been processed using version R4CU4UPD7.4
(for B2~0206+35) and R4CU5UPD2 (for the other two sources) of the
pipeline software, and we followed the ``science threads'' from the
{\it Chandra X-ray Center\/} (CXC) for {\sc CIAO v 1.1.5} to make the
recommended corrections to these data, and in particular to apply the
appropriate gain file, acisD2000-01-29gainN0001.fits.  After screening
out $\leq$ 15 per cent of each observation to avoid intervals of high
background, the exposure times are as given in Table~\ref{xrayobs}.

The radio core positions are known with high precision,
and an overlay of the X-ray and radio images
confirms that the astrometry in the pipeline software used on these
X-ray data sets is imprecise (see http://asc.harvard.edu/mta/ASPECT/).
The X-ray positions given in Table~\ref{xrayobs} have been shifted to
align with the (accurate) radio positions also given in the table.

\begin{table}
\caption{Energy weightings for the PRFs, 0.4 - 5 keV}
\label{prfweights}
\begin{tabular}{lll}
B2 Name &
Energy (keV) &
Weight \\
0206+35 & 0.65 & 0.39 \\
           & 0.9  & 0.24 \\
	   & 1.25 & 0.27 \\
	   & 1.75 & 0.1 \\
0331+39 & 0.7  & 0.47 \\
	   & 1.25 & 0.28 \\
           & 1.75 & 0.13 \\
	   & 3.0  & 0.12 \\
0755+37 & 0.6  & 0.42 \\
	   & 0.9  & 0.19 \\
           & 1.25 & 0.25 \\
 	   & 2.0  & 0.14 \\
\end{tabular}
\end{table}

Since we are interested in features on scales close to {\it
Chandra's\/} spatial resolution, we have found it convenient to
represent the Point Response Function (PRF) with an analytical
function, and we give our fitted parameter values here as they may be
of use to other workers.  The procedure we followed was to use {\sc
ciao/mkpsf} and the CXC-released PRF library to create for each source
an image of the PRF with 0.1~arcsec spatial binning and weighted (in
four energy bins) according to the source spectrum between 0.4 and
5~keV (since these sources are relatively soft).  We convolved the
result with a Gaussian of $\sigma = 0.204$ arcsec to mimic the effects
of aspect smearing (see
http://asc.harvard.edu/ciao1.1/caveats/acis2.html), and then used the
{\sc iraf/imcnts} task to extract the radial profile of the PRF. 
We found the 12-parameter function

$$P(r) = \sum_{i=1}^3 \sum_{j=1}^3 a_{ij} r^{j-1} e^{-{r^2\over 2 s_i^2}}$$

\noindent
with $r$ in units of arcsec, to be a good representation of the PRF.
We first fitted the functional form to a profile extracted from the
low-resolution PRF library, to find the best parameter values for the
PRF wings.  We then fixed the parameters of the broadest Gaussian, and
re-fitted to the radial profile from the high-resolution PRF library
in order to find the best overall representation of the PRF for the
observation.  The energy weightings used for the three sources are
given in Table~\ref{prfweights} and the fitted parameter values are in
Table~\ref{prfpars}.  The resulting profiles are fairly similar, with
half power diameter (HPD) between 0.78 and 0.86 arcsec,
and full width half maximum (FWHM) between 0.53 and 0.62 arcsec. 

\begin{table*}
\caption{PRF Parameter values for the three targets fitted to
$(A_1 + B_1 r + C_1 r^2)e^{-{r^2 \over 2 S_1^2}}+
(A_2 + B_2 r + C_2 r^2)e^{-{r^2 \over 2 S_2^2}}+
(A_3 + B_3 r + C_3 r^2)e^{-{r^2 \over 2 S_3^2}}$
}
\label{prfpars}
\begin{tabular}{lrrrrrrrrrrrr}
B2 Name &
$A_1$ &
$B_1$ &
$C_1$ &
$S_1$ &
$A_2/10^{-2}$ &
$B_2/10^{-2}$ &
$C_2/10^{-3}$ &
$S_2$ &
$A_3/10^{-5}$ &
$B_3/10^{-6}$ &
$C_3/10^{-7}$ &
$S_3$
\\
0206+35 &
1.58 &
$-$1.965 &
1.947 &
0.3523 &
2.059 &
$-$1.361 &
2.618 &
1.407 &
5.146 &
$-$4.695 &
1.247 &
10.68
\\
0331+39 &
2.029 &
$-$3.527 &
1.859 &
0.4326 &
0.9569 &
$-$0.4997 &
0.8484 &
1.638 &
5.456 &
$-$4.871 &
1.256 &
11.36
\\
0755+37 &
1.948 &
$-$3.538 &
3.454 &
0.3462 &
1.902 &
$-$1.219 &
2.272 &
1.452 &
4.987 &
$-$4.538 &
1.211 &
10.79
\\
\end{tabular}
\medskip
\begin{minipage}{\linewidth}
The profiles are normalized to $\sim$~1.03 rather than 1.0; this is
correct for 0.1~arcsec bins and any radial-profile 
extraction routine which works like {\sc iraf/imcnts}
in considering square pixels as falling
entirely in or out of an annulus rather than weighted by area.
\end{minipage}
\end{table*}

\section{X-rays from the galaxy, radio core and jet}

\begin{figure}
\epsfxsize 7.8cm
\epsfbox{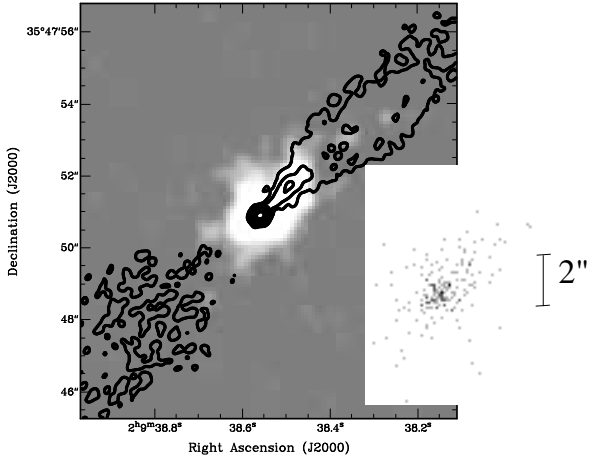}
\caption{
Radio contours on {\it Chandra\/} X-ray image for B2~0206+35. The
radio map is based on VLA archive data at 8.4~GHz
(0\farcs25 FWHM) with contours at
0.24, 0.7, 2, 6, 20, 58 mJy/beam.  The X-ray pixel size is 0.098
arcsec, a Gaussian smoothing with $\sigma$ = 2 pixels has been
applied, and no background is subtracted. Insert, with 2~arcsec scale
bar, shows unsmoothed X-ray counts.
}
\label{0206image}
\end{figure}

\begin{figure}
\epsfxsize 8.5cm
\epsfbox{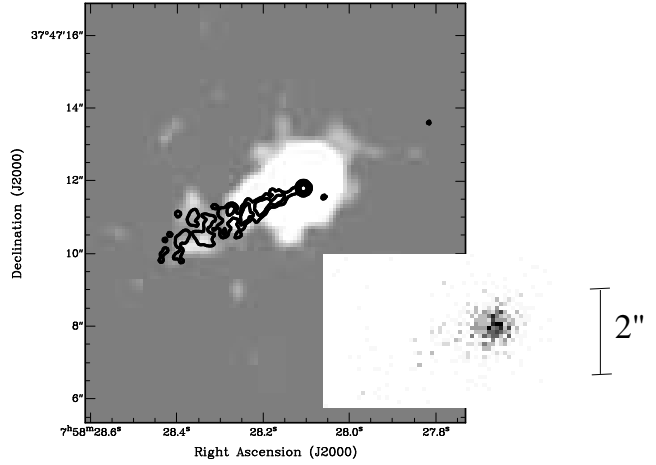}
\caption{
Radio contours on {\it Chandra\/} X-ray image for B2~0755+37. The
radio is Bondi et al.~(2000)'s MERLIN 1.6~GHz data 
(0\farcs15 FWHM) with contours at
0.4, 1.2, 3.6, 10, 40, 300, 900 mJy/beam. The X-ray pixel size is
0.098 arcsec, a Gaussian smoothing with $\sigma$ = 2 pixels has been
applied, and no background is subtracted.  Insert, with 2~arcsec scale
bar, shows unsmoothed X-ray counts.
}
\label{0755image}
\end{figure}

\begin{figure}
\epsfxsize 7.6cm
\epsfbox{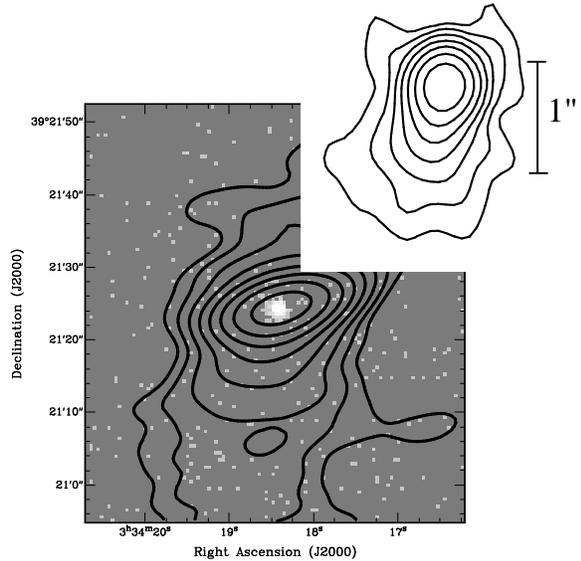}
\caption{Radio contours on {\it Chandra\/} X-ray
image for B2~0331+39.  The radio is Birkinshaw \& Davies (in
preparation)'s VLA map at 4.9~GHz (10\farcs1
$\times$ 3\farcs3 FWHM) with contours at 1.5, 3, 6, 12, 24, 48, 96,
192 mJy/beam.  The X-ray pixel size is 0.493 arcsec, no smoothing is
applied, and no background is subtracted.  The higher-resolution
insert, with 1~arcsec scale bar, shows the inner core at 4.9~GHz:
VLA archive data (0\farcs38 FWHM) with
contours at 1.4, 2.8, 5.6, 11.2, 22.4, 44.8, 89.6 mJy/beam.
}
\label{0331image}
\end{figure}

Compared with the {\it ROSAT\/} PSPC, which preferentially detected
group-scale gas (Worrall \& Birkinshaw 2000), {\it Chandra\/} clearly
sees small-scale emission from B2 radio galaxies.  As seen in Figs
\ref{0206image} and \ref{0755image}, B2~0206+35 and 0755+37 are very
similar, with asymmetric X-ray emission coincident with the dominant
kpc-scale radio jet.  B2~0331+39 (Fig.~\ref{0331image}) shows no such
asymmetry, but in this less powerful radio
source there is no obvious radio jet, either
because of a small viewing angle, which could superimpose the jet
X-ray emission on the core, or because well-collimated kpc-scale
plasma is simply absent in this source. 
The relatively high ratio of core (Table~\ref{radiotab}) to extended 
(Table~\ref{xrayobs}) radio emission in
this source might argue in favour of the former explanation.

\begin{figure}
\epsfxsize 6.5cm
\epsfbox{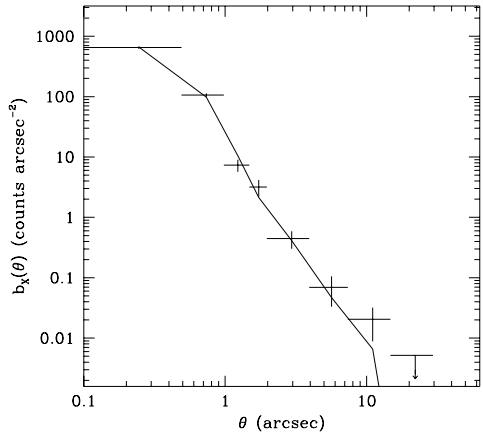}
\caption{
Unlike the other two sources, the background-subtracted X-ray radial
profile of B2~0331+39 fits the PRF extracted for this observation:
$\chi^2 = 7.9$ for 7 degrees of freedom.
}
\label{0331profile}
\end{figure}

The radial profile of B2~0331+39 gives a good fit to the PRF
(Fig.~\ref{0331profile}).  The central emission of the other two
sources is weaker and partially extended: fits to the radial profile
over a semi-circular region in the anti-jet direction in both cases
finds an excess of counts over the PRF.  Good fits
(Table~\ref{xraytab}) are found for a combination of point-like and
extended emission, modelled as a $\beta$~model with counts per unit
area per unit time $\propto (1 +
{\theta^2 \over \theta_{\rm cx}^2})^{0.5 - 3 \beta}$.  Best-fit values
of $\beta$ and $\theta_{\rm cx}$ are 0.7 and 1.15 arcsec, and 0.8 and 2.1
arcsec, for 0206+35 and 0755+37, respectively.  
The extended emission
can be attributed to galaxy-scale gas with a luminosity of a few times
$10^{41}$ ergs s$^{-1}$.  The cooling times of this gas are short
compared with the Hubble time, with mass infall of about one
solar mass per year (Table~\ref{xraytab}).  Upper limits for
B2~0331+39 allow for the presence of galaxy gas at
a luminosity only a factor of a few below that measured in the other two
sources.  It is noteworthy that the galaxy in our sample
without a kpc-scale radio or X-ray jet also has the least
central, X-ray emitting, cooling, gas.

\begin{table}
\caption{Radio components closely matching X-ray emission regions}
\label{radiotab}
\begin{tabular}{lcc@{\hspace{18pt}}clcc}
B2 Name &
$\nu$ & Core &
\multicolumn{4}{c}{Jet} \\
& GHz & $S_{\rm c}$ &  length & radius & $S_{\rm j}$ & $P_{\rm j~min}$\\
&     & mJy & arcsec & arcsec & mJy & $10^{-11}$~Pa\\
0206+35 &8.44 &97$\pm$2& 2 & 0.15 & 20$\pm$2 &3.1\\
0331+39 & 4.86 & 303$\pm$4& -- & -- & -- & -- \\
0755+37 & 1.66 & 150$\pm$4& 4  & 0.2 & 93$\pm$8 &2.1\\
\end{tabular}
\medskip
\begin{minipage}{\linewidth}
For 0331+39, no clear jet is seen; the core flux includes all
the components seen on the high-resolution insert in
Fig.~\ref{0331image}. Jet radius is HWHM of best-fitting Gaussian model.
$P_{\rm j~min}$ is the minimum pressure in the
jet assuming an electron energy spectrum of number index 2.0 between
$\gamma_{\rm min} = 10$ and $\gamma_{\rm max} = 10^5$, no relativistic
protons, and a filling factor of unity.
\end{minipage}
\end{table}

\begin{table*}
\caption{X-ray Components}
\label{xraytab}
\begin{tabular}{lcrr@{\hspace{18pt}}ccc@{\hspace{18pt}}cccc@{\hspace{18pt}}c}
\hline
B2 name &
\multicolumn{3}{c}{Core} &
\multicolumn{3}{c}{Jet} &
\multicolumn{4}{c}{Gas} & $\chi^2$/dof\\
& Counts &$L$ ($10^{41}$ & $S_{1~\rm keV}$ & Counts & $L$ ($10^{41}$ &
$S_{1~\rm keV}$ & Counts &  $L$ ($10^{41}$ &$\dot{M}$&$P_{\rm o}$&\\
& & ergs s$^{-1}$)& (nJy) & & ergs s$^{-1}$) & (nJy) &$r \leq$ 15 kpc
& ergs s$^{-1}$) & M$_\odot$/yr&($10^{-11}$ Pa)&\\
& & & & & & &
 & $r \leq$ 15 kpc &$\theta \leq \theta_{cx}$&&\\
 & \multispan3{\hskip 3pt\hrulefill\hskip 3pt} &
\multispan3{\hskip 3pt\hrulefill\hskip 3pt} & 
\multispan4{\hskip 3pt\hrulefill\hskip 3pt} \\
0206+35 &$76^{+20}_{-40}$&
2.9&9.6&41$\pm14$&
1.6&5.2&$90^{+27}_{-32}$&
2.6&$0.7_{-0.45}^{+1.0}$&$8.5_{-5.5}^{+19.5}$&
0.1/4\\
0331+39 &756$\pm$28&
16.0&200.0&--&--&--&
$< 48$& $<$0.75&--& --&
7.9/7\\
0755+37 &$479^{+22}_{-80}$&
19.0&47.6&98$\pm26$&
3.9&9.7&$109^{+67}_{-23}$&
4.2&$0.25_{-0.17}^{+0.24}$&$3_{-1.5}^{+8}$
&6.8/6\\
\hline
\end{tabular}
\vbox{\vskip 5pt
\begin{minipage}{\linewidth}
Counts and luminosities are for energies between 0.4 and 5
keV.
$\chi^2$/dof is for spatial fit to a point-source for 0331+39, and to
a point source plus an extended component in the anti-jet semicircle
for 0206+35 and 0755+37, although the counts and luminosities given
are over all angles.  Mass deposition rate,
$\dot{M}$, and
central gas pressure, $P_{\rm o}$, are calculated (Birkinshaw \&
Worrall 1993) over all acceptable
combinations of $\beta$ and $\theta_{\rm cx}$.  The $3\sigma$ upper
limits for 0331+39 are based on a gas model with
$\beta = 0.75$, $\theta_{\rm cx} = 1.6$ arcsec,
$kT = 0.5$~keV, and 0.3~solar abundance.  Parameter
values other than counts use the best-fit spectral parameters from
Table~\ref{spectraltab}, with the power-law parameters applied to
both core and jet.
\end{minipage}}
\end{table*}

\begin{table*}
\caption{X-ray Spectra}
\label{spectraltab}
\begin{tabular}{lccccc}
B2 Name & Power law energy & $N_H$ intrinsic & $kT$ & abundance & $\chi^2$/dof \\
         & index, $\alpha$  & ($10^{20}$ cm$^{-2}$) & (keV) & /solar    &\\
0206+35 & $1.05_{-0.85}^{+0.57}$ & -- & 
$0.56_{-0.11}^{+0.08}$ & $0.06_{-0.04}^{+0.9}$ & 8.4/5\\
0331+39 & $1.62 \pm 0.16$ & $6.7 \pm 3.3$ &
 -- & -- & 29.5/30\\
0755+37 & $1.1_{-0.4}^{+0.1}$ & $ < 10.4 $ & 
$0.84_{-0.16}^{+0.07}$ & $0.8_{-0.74}^{+0.1}$ & 19.0/26\\
\end{tabular}
\medskip
\begin{minipage}{\linewidth}
Absorption through the Galactic column (Table~\ref{xrayobs}) 
has been applied to all fits.
Intrinsic column density is applied only to the power law components:
it is neither required nor usefully constrained in the fits for 0206+35.
Errors are $1\sigma$ for one interesting parameter. Upper limits are
$3\sigma$ significance.
\end{minipage}
\end{table*}

The X-ray spectral fits to the net counts within a radius of 10 arcsec
around each source (Table~\ref{spectraltab}) support the spatial
modelling. While B2~0331+39 gives an acceptable fit to a power law,
the other two sources either give unacceptable fits to
single-component models (power law or Raymond-Smith thermal), or
$\chi^2$ improves significantly when the second component is added.
The normalizations of the power-law and thermal models in the
two-component fits imply a division of counts between core$+$jet and gas
which agrees well with the spatial separation of components.  The gas
has a temperature reasonable for its luminosity as compared
with the extrapolation of the temperature-luminosity relationship
for the larger-scale group/cluster atmospheres around low-power
radio galaxies (see figure~7 of Worrall \& Birkinshaw 2001).

Our {\it ROSAT\/} studies of low-power radio galaxies had separated
some of the PSPC emission into thermal gas associated with the group
or cluster (Worrall \& Birkinshaw 1994, 2000), and attributed the
remaining compact emission, or that unresolved with the HRI (Canosa et
al.~1999), to a component associated with the small-scale radio
structures.  Although {\it Chandra\/} resolves a fraction of the
emission we previously associated with the AGN (compare table 4 of
Canosa et al.~2000 with Table~\ref{xraytab}), our description of a
low-power radio galaxy as emitting significant small-scale
radio-related X-ray emission is confirmed. {\it
Chandra's\/} superior spatial resolution reveals power-law emission
from an unresolved core in all three sources, and X-ray jets
in the two sources with small-scale radio jets.

Although three is too small a sample to claim a correlation, it is
interesting that the core radio flux densities (which are assumed
typical in having flat radio spectra) are in the same rank order as
the core X-ray flux densities.  The core X-ray spectra are soft.  A
small intrinsic absorption of a few 10$^{20}$ cm$^{-2}$ is measured in
B2~0331+39,
and an upper limit for
B2~0755+37 allows for comparable amounts.  Such absorption is
consistent with expectations from the dusty medium which {\it HST\/}
finds to be typical in such radio galaxies: compare the detailed
analysis of Ferrarese \& Ford (1999) for one source with the general
findings concerning dust in B2 radio galaxies of Capetti et
al.~(2000).  It remains plausible that the core X-ray emission is
associated with the relativistic particles directly responsible for
VLBI-scale radio emission, as our earlier work suggested (Worrall \&
Birkinshaw 1994).

The galaxy-scale gas may play a role in the confinement, disruption,
or slowing of the jets in the crucial region within a few kpc of the
nucleus.  Calculations of internal jet pressures are model dependent.
In Table~\ref{radiotab} we give values for the minimum pressure on the
simplest assumption that the jets are in the plane of the sky and that
relativistic effects can be ignored.  The effects of relativistic
beaming and projection will tend to reduce these values, by a factor
of about four in the case of Bondi et al. (2000)'s model for
B2~0755+35. It is interesting that the kpc-scale jet pressures are
similar to the central gas pressures.  Deeper {\it Chandra\/} X-ray
observations are needed for detailed modelling of the interface
between the gas and jet plasma, such as in our study of
B2~0104+32 (3C~31: Hardcastle et al., in preparation).

\section{Discussion and Conclusions}

The {\it Chandra\/} observations of B2 radio galaxies support the
general picture of a low-power radio galaxy having galaxy-scale gas, a
compact X-ray core, and X-ray jet emission associated with fast,
one-sided, kpc-scale, radio jets.  The presence also of group- or
cluster-scale gas is known from {\it ROSAT\/}, but this appears as
structureless background in our high-resolution, reduced
field-of-view, {\it Chandra\/} images.

For both sources with radio and X-ray jet emission, the jet emission
is an appreciable fraction of the core emission (of order 20-50 per
cent) in both the X-ray and radio.  This similarity of X-ray to radio
ratio for the jets and cores argues in favour of the core and jet
X-ray components both being radio-related in origin, supporting our
earlier conclusions (Worrall \& Birkinshaw
1994).

Contending emission mechanisms for the jet X-rays are synchrotron
radiation from particles of higher energy than those contributing the
radio emission, or Compton scattering of jet-produced or external
photons.  The X-ray spectrum can help to decide between these
possibilities, since it has a direct relationship to the spectrum of
the particles responsible for the radiation.  A steep X-ray spectrum
would tend to rule out a Compton-scattering origin, and instead
suggest synchrotron emission from high-energy electrons of steeper
spectrum than those emitting in the radio.  In B2~0331+39 the X-ray
spectrum is indeed steep, but these shallow X-ray observations lead to
constraints that are too poor to draw useful conclusions for
B2~0206+35 and 0755+37.  However, the X-ray flux levels of the jets
are most easily explained by a synchrotron model. Synchrotron
self-Compton emission with an equipartition magnetic field produces an
X-ray flux density $\sim 3$ orders of magnitude below the observed
value, using the jet parameters given in Table~\ref{radiotab}.  FRI
jets are generally thought to move too slowly (\S1) for the effective
energy density in microwave-background photons to be significantly
increased by boosting (Tavecchio et al.~2000), and, even if the regions
concerned do have very high flow velocities, the jets are unlikely to
be close enough to the line of sight for the mechanism to be
important. The dominant photon
population in the jets is likely to be beamed radiation from the
active nuclei, and although estimates of the inverse Compton X-ray
yield are strongly model dependent, we find they are likely to fall
short by an order of magnitude under plausible assumptions
(Hardcastle, Birkinshaw \& Worrall~2001b).  A synchrotron model
requires a steepening of the spectral index between the radio and
X-ray if it is not to over-produce the X-rays, but such spectral
breaks are observed in well-studied optical synchrotron jets such as
M87 and 3C~66B, for which multiwavelength data are compiled in
Hardcastle et al.~(2001b).

It is easier to detect jet X-ray emission with {\it Chandra\/} than
jet optical emission with {\it HST\/},
at least for sources with relatively high core dominances such as these.  
In two out of our three sample
sources, and both of those with kpc-scale radio jets, X-ray jets are
detected.  In contrast, Capetti et al.~(2000) report detection of only
three of 57 representative B2 radio galaxies observed in an {\it
HST\/} snapshot survey, even though roughly 45 per cent of all B2
radio galaxies have kpc-scale radio jets (Parma et al.~1987).
Interestingly, B2~0755+37 is one of the three sources with detected
optical jets, and so more detailed modelling of the spectral energy
distribution of this jet will be possible.

We are currently analysing data for five other low-power radio
galaxies that we have observed more recently with {\it Chandra\/}, two
of which are from the B2 sample, and all five show jet features.  At
the spatial resolution possible with {\it Chandra\/}, X-ray jets are
indeed common.  As more data are forthcoming it should be possible to
make a detailed study of the underlying physics which controls the
strength of X-ray emission in these low-power jets.

\section*{Acknowledgments}

We thank Marco Bondi for the processed MERLIN image of B2~0755+37, 
John Biretta and Paola Parma for
use of their VLA archival data on B2~0206+35 and
B2~0331+39, respectively, and
staff of the CXC for help concerning calibrations and the {\sc ciao}
and {\sc ds9} software.

\label{lastpage}

\end{document}